\documentclass[apj]{emulateapj}

\begin{document}

\shorttitle{X-rays from PSR J0437--4715}
\shortauthors{Bogdanov}

 \title{The Nearest Millisecond Pulsar Revisited with \textit{XMM-Newton}: \\ Improved Mass-Radius Constraints for PSR J0437--4715} 

\author{Slavko Bogdanov} 

\affil{Columbia Astrophysics Laboratory, Columbia University, 550 West 120th Street, New York, NY 10027, USA;\\ and Department of Physics, McGill University, 3600 University Street, Montreal, QC H3A 2T8, Canada;\\ slavko@astro.columbia.edu}

\begin{abstract}
  I present an analysis of the deepest X-ray exposure of a radio
  millisecond pulsar (MSP) to date, an \textit{X-ray Multi
    Mirror-Newton} European Photon Imaging Camera spectroscopic and
  timing observation of the nearest known MSP, PSR J0437--4715. The
  timing data clearly reveal a secondary broad X-ray pulse offset from
  the main pulse by $\sim$0.55 in rotational phase. In the context of
  a model of surface thermal emission from the hot polar caps of the
  neutron star, this can be plausibly explained by a magnetic dipole
  field that is significantly displaced from the stellar center. Such
  an offset, if commonplace in MSPs, has important implications for
  studies of the pulsar population, high energy pulsed emission, and
  the pulsar contribution to cosmic ray positrons.  The continuum
  emission shows evidence for at least three thermal components, with
  the hottest radiation most likely originating from the hot
  magnetic polar caps and the cooler emission from the bulk of the
  surface.  I present pulse phase-resolved X-ray spectroscopy of PSR
  J0437--4715, which for the first time, properly accounts for the
  system geometry of a radio pulsar. Such an approach is essential for
  unbiased measurements of the temperatures and emission areas of
  polar cap radiation from pulsars.  Detailed modelling of the thermal
  pulses, including relativistic and atmospheric effects, provides a
  constraint on the redshift-corrected neutron star radius of $R>11.1$
  km (at 3$\sigma$ conf.) for the current radio timing mass
  measurement of $1.76$ M$_{\odot}$. This limit favors ``stiff''
  equations of state.
\end{abstract}

\keywords{pulsars: general --- pulsars: individual (PSR J0437--4715)
--- stars: neutron --- X-rays: stars --- relativity}

\section{INTRODUCTION}
PSR J0437--4715 was discovered by \citet{John93} in the Parkes
southern radio pulsar survey. At a distance of $156.3\pm1.3$ pc
\citep{Del08}, it is the nearest known rotation-powered ``recycled''
millisecond pulsar (MSPs). It has properties typical of the Galactic
population of MSPs, with a spin period $P=5.76$ ms and spin-down rate
(after kinematic corrections) of $\dot{P}\equiv {\rm d}P/{\rm d}t =
1.0\times10^{-20}$ s s$^{-1}$, corresponding to a dipole
magnetic field strength $B=3\times10^8$ G, a characteristic age
$\tau\approx4.9$ Gyr, and spin-down luminosity
$\dot{E}=3.8\times10^{33}$ ergs s$^{-1}$. The pulsar is bound to a
$0.2$ M$_{\odot}$ white dwarf companion in a 5.74 day circular orbit
\citep{Bailyn93}.

In soft X-rays, PSR J0437--4715 was the first MSP to be firmly
detected, in a serendipitous discovery in the course of the
\textit{ROSAT} all sky survey \citep{Beck93}. Subsequent observations
with \textit{ROSAT} \citep{Beck99}, the \textit{Extreme Ultraviolet
  Explorer} Deep Survey Instrument \citep{Hal96}, \textit{Chandra}
ACIS-S and HRC-S \citep{Zavlin02}, and \textit{XMM-Newton} EPIC MOS
and pn \citep{Zavlin06} revealed a single broad, asymmetric pulse and
a relatively soft spectrum, with at least two distinct spectral
components.  The bulk of emission up to $\sim$2 keV is thermal in
nature. The small inferred effective radii ($\lesssim$2 km) suggest
that this radiation is associated with the neutron star polar
caps. Given the large characteristic age of the pulsar, the heat is
most likely not due to neutron star cooling but is instead
continuously supplied by an energetic return current from the
magnetosphere above the polar caps \citep[see, e.g.,][]{Hard02}. Above
$\sim$2 keV, \citet{Zavlin02} have identified an additional faint
component that is best described by a power-law. The nature of this
spectral tail is less certain as it could be due to a variety of
plausible mechanisms such as pulsed non-thermal radiation due to
particle acceleration in the pulsar magnetosphere, unpulsed
synchrotron radiation due to an unresolved, faint compact pulsar wind
nebula or intrabinary shock, or a spectral tail caused by inverse
Compton scattering of the soft thermal photons by particles (most
likely electrons and positrons) of small optical depth \citep{Bog06}.
PSR J0437--4715 has also been detected in the UV \citep{Kar04,Dur12}
and recently identified as a pulsed source in $\gamma$-rays ($>$100
MeV) with the \textit{Fermi} Large Area Telescope \citep{Abdo09}.

%
%   FIGURE 1
%
\begin{figure*}[!t]
\begin{center}
\includegraphics[width=0.55\textwidth]{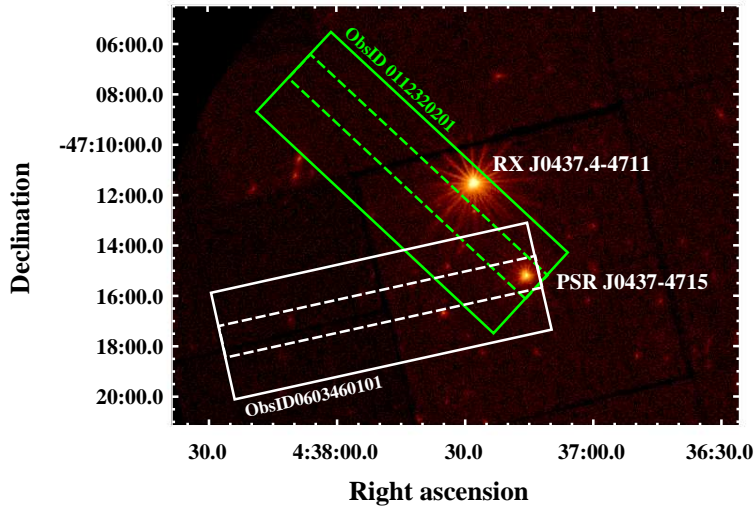}
\end{center}
\caption{\textit{XMM-Newton} MOS1 and MOS2 mosaic image of the field
  around PSR J0437--4715 in the 0.1--10 keV band. The rectangles show
  the relative positions and orientations of the active EPIC pn chip 4
  used in observations 0603460101 (\textit{white}) and 0112320201
  (\textit{green}), with the timing readout direction indicated by the
  arrow. The dashed lines delineate the EPIC pn detector columns
  used to extract the source counts from the pulsar. Note the
  significant contribution of photons from the background source RX
  J0437.4--4711 in the source extraction columns of observation
  0112320201.}
\end{figure*} 
Based on existing X-ray observations, modeling of the pulsed emission
from PSR J0437--4715 has revealed unique insight into the properties
of the stellar surface, magnetic field configuration, and neutron star
structure. For instance, \citet{Pavlov97}, \citet{Zavlin98}, and
\citet{Bog07} have demonstrated that a hydrogen atmosphere is likely
present at the neutron star surface, as expected for a recycled pulsar
\citep{Alp82}. The morphology of the thermal pulsations seen in the
archival \textit{XMM-Newton} data point to an offset of the magnetic
axis from the neutron star center \citep{Bog07}.  In addition, the
compactness of PSR J0437--4715 is found to be $R/R_S>1.6$ (where
$R_S\equiv2GM/c^2$ is the Schwarzschild radius) at 99.9\% confidence,
which for the best available mass measurement \citep[$1.76\pm0.20$
  M$_{\odot}$][]{Verb08} corresponds to $R\gtrsim8.3$ km
\citep{Bog07}. As shown in \citet{Bog08}, in principle, with deeper
X-ray exposures it may be possible to obtain even tighter limits on
the allowed neutron star equations of state. As the nearest and
brightest known MSP, PSR J0437--4715 is the best-suited target for
detailed investigation, especially given the availability of an
independent mass measurement derived from radio timing
observations. This combination can potentially provide stringent
constraints on neutron star structure.

In this present paper, I report on a new \textit{XMM-Newton}
spectroscopic and timing observation of PSR J0437--4715, which sheds
new light on the X-ray properties of MSPs and the physics of neutron
stars, in general.  The paper is organized as follows. In \S 2, I
overview the observations and data reduction. In \S3, I describe the
characteristics of the pulse profile, In \S4 and \S5, I present
phase-averaged and phase-resolved spectroscopic analyses of the
pulsar. In \S 6, I present modelling the pulsed emission from PSR
J0437--4715. I offer conclusions in \S 7.

\section{OBSERVATION AND DATA REDUCTION}
PSR J0437--4715 was revisited with \textit{XMM-Newton} between 2009
December 15 and 17 (observation ID 0603460101) in a continuous
130-kilosecond exposure, corresponding to the entire available
observing time of revolution 1835.  The European Photon Imaging Camera
(EPIC) MOS1/2 instruments were configured for full imaging mode. The
EPIC pn was used in fast timing mode, in which only detector chip 4 is
active, enabling 30 $\mu$s relative timing precision while sacrificing
one imaging dimension. For the MOS1/2 and pn, the thin optical filter
was in place. The dispersed Reflection Grating Spectrometer (RGS) data
yielded no interesting spectral or timing information, owing to the
relatively faint nature of the pulsar, and were thus not used in this
investigation.

The data reduction, imaging, and timing analyses were carried out
using SAS\footnote{The \textit{XMM-Newton} SAS is developed and
  maintained by the Science Operations Centre at the European Space
  Astronomy Centre and the Survey Science Centre at the University of
  Leicester.} 10.0.0 and FTOOLS\footnote{Available for download at
  http://heasarc.gsfc.nasa.gov/ftools/} 6.10, while the spectral
analysis was conducted in XSPEC\footnote{See
  http://heasarc.nasa.gov/docs/xanadu/xspec/} 12.6.0q.  The raw (ODF)
MOS and pn datasets was reprocessed with the SAS {\tt emchain} and
{\tt epchain} tools, respectively, and filtered for instances of
strong background flares.  Discarding time intervals with severely
contaminated data results in 104.9, 104.9, and 120.4 ks of usable
exposure time for MOS1, MOS2, and pn, respectively. Subsequently, the
recommended processing filters (flag, pulse invariant, and pattern)
were applied to generate the data used in the spectral and pulse
profile analyses.

For the phase-integrated spectroscopy, the counts from the pulsar in
the MOS 1/2 data were extracted from 60'' circles centered on the
source position derived from radio timing \citep{Verb08}, which
contain $\sim$90\% of the total energy at $\sim$1.5 keV.  The
background was estimated by considering three source-free regions in
the immediate vicinity of the pulsar. For the phase-resolved
spectroscopy and pulse profile modelling of the EPIC pn dataset, the
counts from the MSP were obtained from RAWX columns 29--43
(inclusive), corresponding to a 30.75$\arcsec$ circle in full imaging
mode, which encloses $\sim$88\% of the total energy for 0.3--2
keV. This narrow extraction region was chosen so as to minimize the
large background level (owing to the one-dimensional imaging mode used
for the pn detector), which dominates beyond $\sim$30'' from the
source position. To obtain a reliable estimate, the background was
taken from source-free readout columns on both sides of the columns
containing the pulsar.  For the spectroscopic analysis of the EPIC MOS1/2
and pn data, the extracted counts were grouped using at least 30 and
150 counts per bin, respectively.

The archival \textit{XMM-Newton} observation of PSR J0437--4715
(observation ID 0112320201, revolution 519), with the same
instrument configuration as the new observation, was performed with an
unfavorable mean telescope position angle (137$^{\circ}$). As a
consequence, the readout columns containing the pulsar emission are
significantly contaminated by the background active galactic nucleus
RX J0437.4--4711 \citep{Hal96b}. In contrast, the new observation was
carried out at a position angle that avoids the AGN entirely
(192$^{\circ}$, see Figure 1) thus resulting in significant
improvement in the quality of the timing data and a more reliable
estimate of the background in the EPIC pn. In light of the problems
with the archival data, the two EPIC pn observations were not combined
and only the new, uncontaminated dataset was used.
%
%    FIGURE 2
%
\begin{figure}[t!]
\begin{center}
\includegraphics[width=0.45\textwidth]{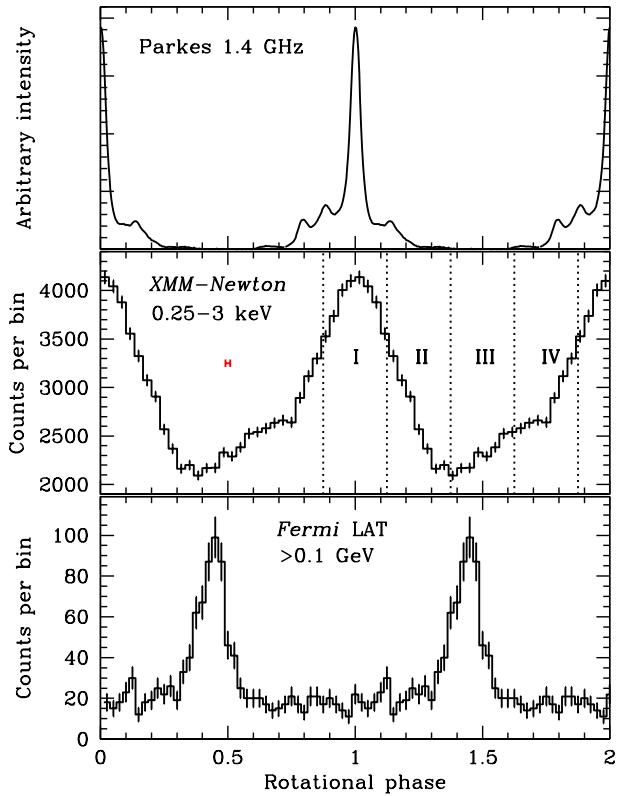}
\end{center}
\caption{A comparison between the \textit{XMM-Newton} EPIC pn X-ray
  pulse profile in the 0.3--2 keV range (\textit{middle panel}), the
  template profile at 1.4 GHz (\textit{top panel}), and the
  \textit{Fermi} LAT profile (\textit{bottom panel}) of PSR
  J0437--4715. Phase zero is defined based on the radio ephemeris. The
  error bar in the middle panel corresponds to the 70 $\mu$s absolute
  timing uncertainty of \textit{XMM-Newton}. The vertical dotted lines
  mark the phase intervals used in the phase-resolved
  spectroscopic analysis discussed in \S5.}
\end{figure}

\section{PULSE PROFILE ANALYSIS}
To obtain the X-ray pulse profile of PSR J0437--4715, the arrival
times of the photons extracted from the pn source region (Fig.~1) were
translated to the solar system barycenter with the {\tt barycen} tool
in SAS, assuming the pulsar position derived from radio timing
\citep{Verb08} and the JPL DE405 solar system ephemeris. The corrected
arrival times were folded at the radio ephemeris of PSR J0437--4715
from \citet{Verb08} using the TEMPO2\footnote{Available for download
  at http://www.atnf.csiro.au/research/pulsar/tempo/} pulsar timing
package.  For an illustrative comparison, Figure 2 shows the radio,
X-ray and $\gamma$-ray pulse profiles of PSR J0437--4715 all folded
using the same ephemeris, hence showing the absolute phase alignment
of the three pulse profiles.

Previous observations have uncovered that the X-ray pulsations from
PSR J0437--4715 is characterized by a single broad and asymmetric peak
that is significantly wider than the radio and $\gamma$-ray
counterparts \citep{Zavlin02,Zavlin06}.  The substantial improvement
in photon statistics of the new data clearly reveals a previously
unseen ``hump'' at phases 0.5--0.75.  For surface polar cap radiation,
this feature likely corresponds to the secondary hot spot on the
``far'' side of the neutron star. The implied peak-to-peak separation
of the two pulses in phase is $\sim$0.55 as measured from the more
prominent to the fainter pulse, substantially different from 0.5 as
expected from antipodal polar caps. This implies hot spots that are
not diametrically opposite, presumably due to a significant offset of
the magnetic dipole axis from the center of the star.  Note that for
PSR J0437--4715, the magnitude of the Doppler effect that would be
induced by the rapid stellar rotation is not large enough to cause the
apparent asymmetry, especially when a H atmosphere is assumed
\citep[see Figure 1 in][]{Bog07}.
%Moreover, this
%effect would produce an asymmetry in the opposite sense
%\citep{Bra00,Pou03,Pou06}.

The pulsed fraction (defined by convention as the portion of counts
above the pulse minimum) of the background-subtracted lightcurves was
determined from the pulse profile modelling presented in \S6 in five
energy bands: 0.275--0.35, 0.35--0.55, 0.55--0.75, 0.75--1.1, and
1.1--1.7 keV. This approach yields $32\%\pm1\%$, $35\%\pm1\%$,
$37\%\pm1\%$, $37\%\pm1\%$, and $35\%\pm2\%$, respectively. It has
been shown that for the range of plausible NS masses and radii,
isotropic surface thermal emission from a neutron star surface, such
as that due to a blackbody, cannot exceed $\sim$33\% even from
point-like hot spots for the plausible range of NS masses and radii
\citep{Psa00,Belo02}. Nonetheless, the computed pulsed fractions are
still in agreement with a predominantly thermal origin of the observed
radiation if a H atmosphere is present on the neutron star surface. This
is because the inherently anisotropic pattern of the emergent
radiation from such an atmosphere can produce much larger
rotation-induced modulations \citep[see, e.g.,][]{Zavlin96,Zavlin98}.
The slightly lower pulsed fraction at energies below $\sim$$0.35$ keV
can be attributed to the contribution of emission from a larger
portion of the stellar surface, as suggested by the UV emission from
the pulsar \citep{Dur12} and the spectroscopic analysis in \S4, which
would be less modulated due to its greater physical extent.  The pulse
profiles for the other four energy bands are consistent with having
identical pulsed fractions. Although the number of source photons is
quite sparse above 2 keV, faint modulations similar to those at lower
energies are seen in the 2--6 keV band. Closer inspection of the model
spectrum shown in Figures 3, however, reveals a significant
contribution from the hot thermal component, implying that the
modulation may still be due to thermal emission. Unfortunately, due to
the high background of the EPIC pn data, combined with the significant
decline in the telescope effective area at higher energies, no useful
pulse profile information is available above $\sim$3 keV, where the
contribution from the thermal components is negligible. Thus, only an
uninteresting limit of $\lesssim$90\% pulsed fraction for the
background-subtracted emission can be placed, precluding any insight
into the origin of the power-law tail.

%
%  FIGURE 3
%
\begin{figure}[t!]
\begin{center}
\includegraphics[width=0.45\textwidth]{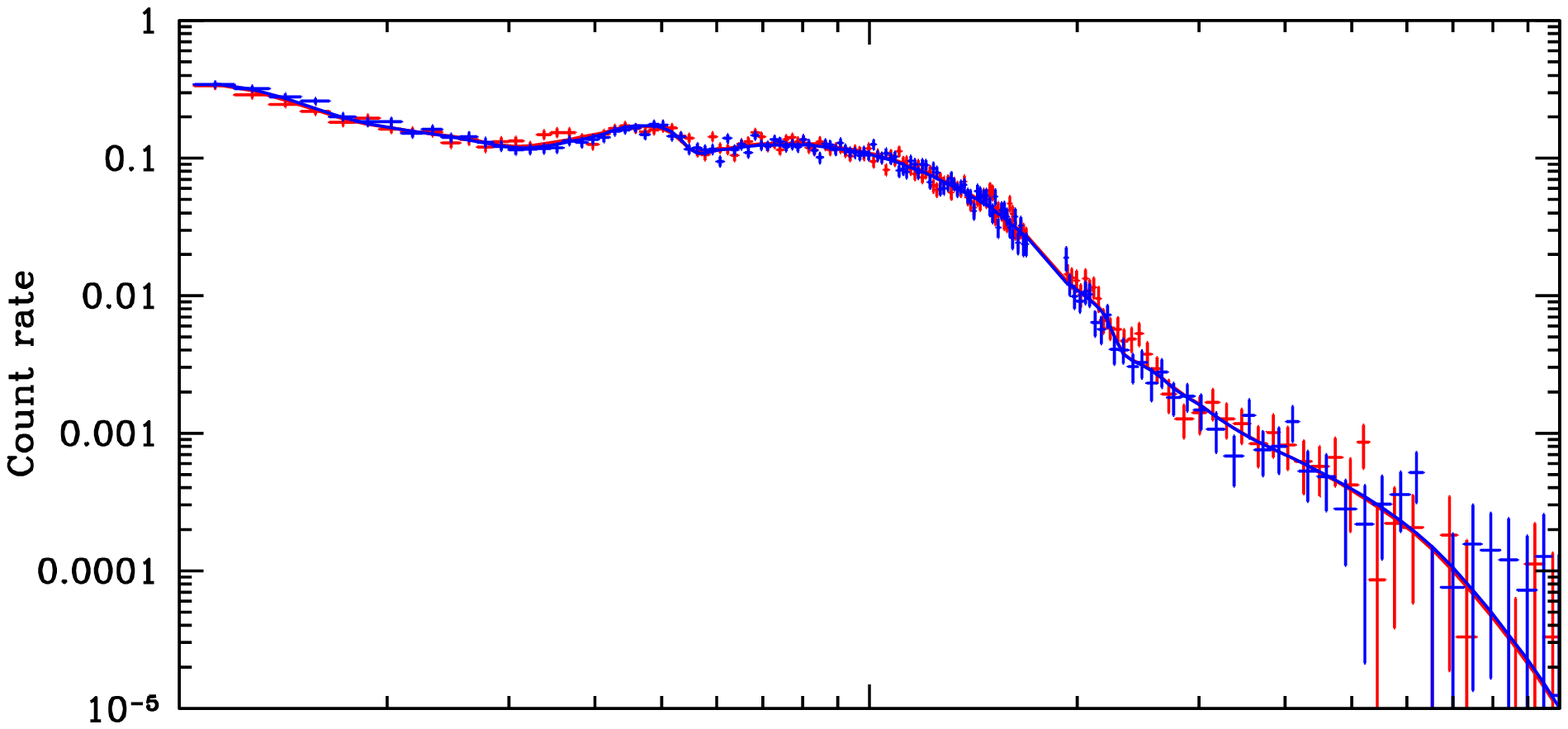}\\
\includegraphics[width=0.45\textwidth]{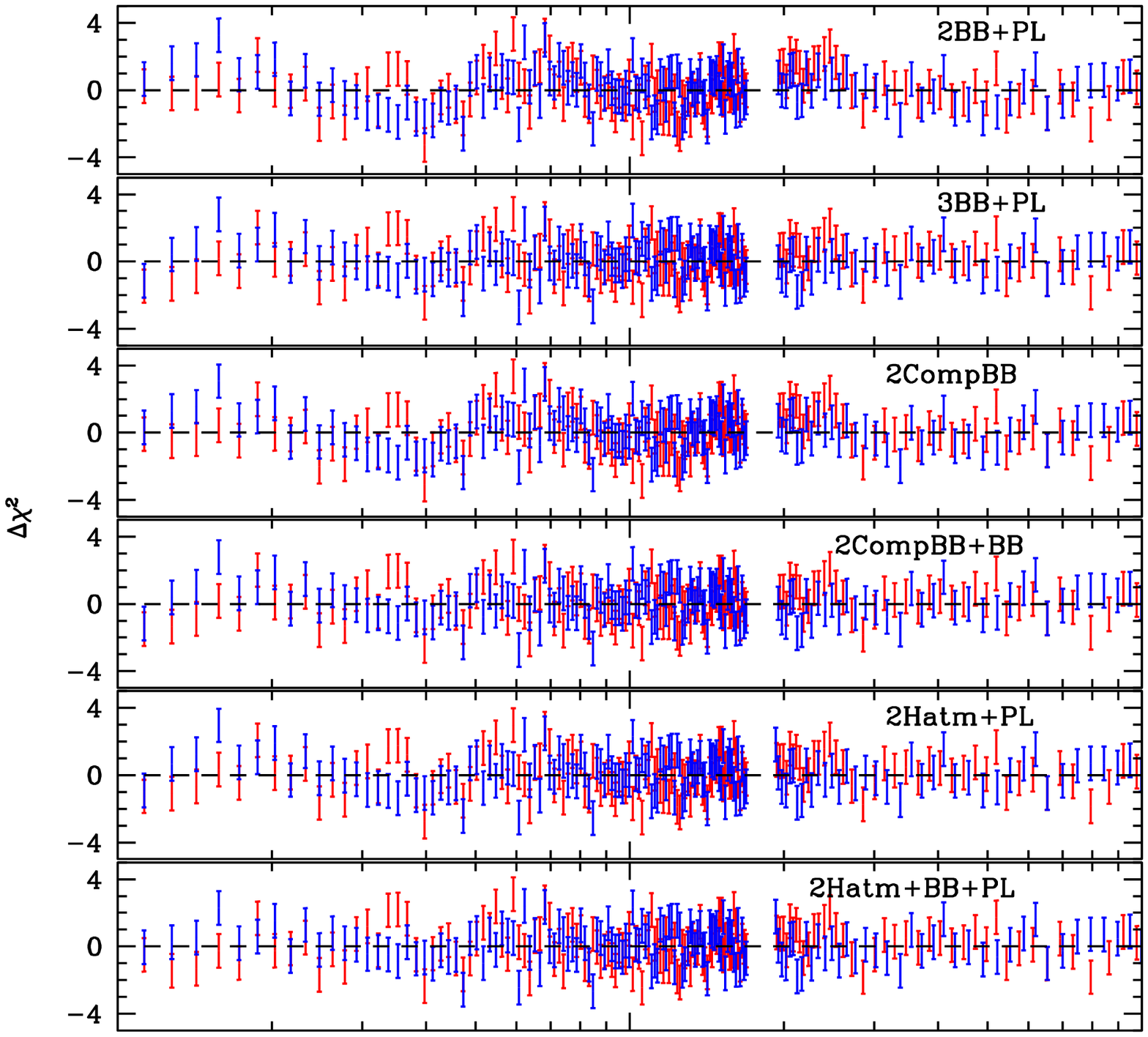}\\
\includegraphics[width=0.45\textwidth]{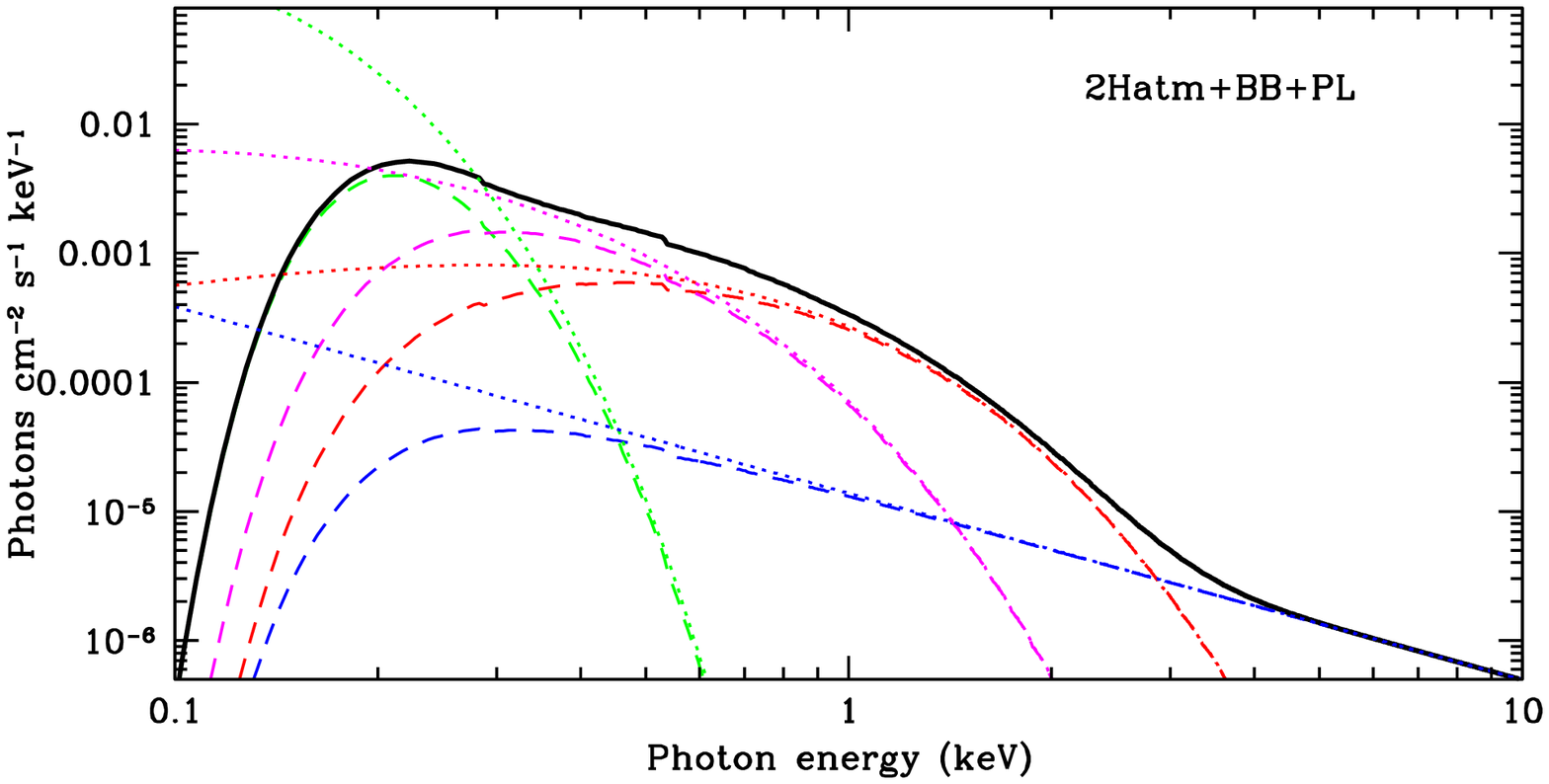}
\caption{The \textit{XMM-Newton} X-ray EPIC MOS1/2 spectra of PSR
  J0437--4715 fitted with a two-temperature H atmosphere model plus a
  power-law and a cool blackbody (\textit{solid line}). The middle six
  panels show the fit residuals for the various models listed in Table
  1. The bottom panel shows the best fit two-temperature H atmosphere
  model plus a power-law and a cool blackbody.  The dashed and dotted
  lines show the individual absorbed and unabsorbed model components,
  respectively, while the solid line shows the total spectrum. See text
  and Table 1 for best fit parameters and acronym definitions.}
\end{center}
\end{figure}

\begin{deluxetable*}{lcccccccccl}
%\rotate
\tabletypesize{\footnotesize} 
\tablecolumns{11} 
\tablewidth{0pt}
\tablecaption{\textit{XMM-Newton} EPIC MOS1/2 best-fit spectral models and unabsorbed fluxes for PSR J0437--4715}
\tablehead{ \colhead{}              & 
            \colhead{$N_{\rm H}$}     &
            \colhead{$T_{\rm eff,1}$} & 
            \colhead{$R_{\rm eff,1}$\tablenotemark{b}} & 
            \colhead{$T_{\rm eff,2}$} & 
            \colhead{$R_{\rm eff,2}$\tablenotemark{b}} &  
            \colhead{ } & 
            \colhead{ } & 
            \colhead{$\Gamma$} &  
            \colhead{$F_{\rm X}$\tablenotemark{c}} & 
            \colhead{$\chi^2_{\nu}$/dof} \\
            \colhead{Model\tablenotemark{a}} & 
            \colhead{($10^{19}$ cm$^{-2}$)} & 
            \colhead{($10^6$ K)} & 
            \colhead{(km)} & 
            \colhead{($10^6$ K)} & 
            \colhead{(km)} &   
            \colhead{ }  & 
            \colhead{ } &  
            \colhead{ } & 
            \colhead{(0.1--10 keV)}  & 
            \colhead{ } }

\startdata
BB$(\times$2)+PL	&	$5.7^{+1.7}_{-1.5}$	&	$3.02^{+0.14}_{-0.12}$	&	$0.05^{+0.03}_{-0.02}$	&	$1.70^{+0.18}_{-0.16}$	&	$0.11^{+0.07}_{-0.06}$	& - & - &	$2.78^{+0.04}_{-0.07}$ &	$1.80(1.13)$	&	$1.13/282$	\\
	&	$<$$0.09$	&	$2.76^{+0.03}_{-0.05}$	&	$0.07^{+0.02}_{-0.01}$	&	$0.99^{+0.01}_{-0.04}$	&	$0.46^{+0.17}_{-0.11}$	&	- &   -	& $1.56$ &	$1.34(0.20)$	&	$1.48/283$	\\
	&	$7$	&  $3.03^{+0.14}_{-0.12}$	&	$0.04^{+0.02}_{-0.02}$	&	$1.77^{+0.15}_{-0.13}$ &   $0.10^{+0.05}_{-0.04}$ & -	& - & $2.81^{+0.03}_{-0.03}$ &	$1.88(1.25)$	&	$1.13/283$	\\
Hatm($\times$2)+PL	&	$4.5^{+2.6}_{-2.2}$	&	$2.14^{+0.16}_{-0.11}$	&	$0.15^{+0.10}_{-0.06}$	&	$0.75^{+0.08}_{-0.08}$	&	$1.15^{+1.10}_{-0.78}$	&	- &  -	& $2.70^{+0.10}_{-0.24}$  &	$1.66(0.90)$	&	$1.15/282$	\\
	&	$0.87^{+0.59}_{-0.55}$	&	$2.22^{+0.05}_{-0.04}$	&	$0.18^{+0.04}_{-0.05}$	&	$0.55^{+0.02}_{-0.01}$	&	$5.1^{+2.5}_{-3.5}$	& - &  -	& $1.56$  &	$1.41(0.15)$	&	$1.14/283$	\\
	&	$7$	&	$2.03^{+0.12}_{-0.11}$	&	$0.17^{+0.05}_{-0.05}$	&	$0.75^{+0.08}_{-0.08}$	&	$0.65^{+0.19}_{-0.25}$	&	- &  -	& $2.80^{+0.04}_{-0.03}$  &	$1.85(1.20)$	&	$1.15/283$	\\

\hline
\colhead{} &  &
\colhead{} & \colhead{} & \colhead{} & \colhead{} &  \colhead{$T_{\rm eff,3}$} & \colhead{$R_{\rm eff,3}$}  &  \colhead{}   & \colhead{} & \colhead{} \\
\colhead{} &  &
\colhead{} & \colhead{} & \colhead{} & \colhead{} &  \colhead{($10^6$ K)} & \colhead{(km)}  & \colhead{}  & \colhead{} & \colhead{} \\
\hline
BB($\times$3)+PL	&	$18.2^{+4.8}_{-4.0}$	&	$3.16^{+0.11}_{-0.10}$	&	$0.05^{+0.02}_{-0.01}$	&	$1.51^{+0.08}_{-0.06}$	&	$0.21^{+0.11}_{-0.10}$	&	$0.44^{+0.06}_{-0.04}$	&	$4.6^{+4.8}_{-3.7}$	& $1.78^{+0.35}_{-0.17}$	&	$2.65(0.19)$	& $1.08/280$		\\
	&	$16.7^{+5.5}_{-3.6}$	&	$3.21^{+0.09}_{-0.09}$	&	$0.04^{+0.02}_{-0.01}$	&	$1.54^{+0.06}_{-0.08}$	&	$0.21^{+0.12}_{-0.08}$	&	$0.46^{+0.02}_{-0.04}$	&	$4.0^{+3.5}_{-3.0}$	& $1.56$	&	$2.46(0.16)$	& $1.07/281$		\\
    &	$7$ & $3.20^{+0.08}_{-0.02}$	&	$0.04^{+0.02}_{-0.01}$	& $1.57^{+0.09}_{-0.06}$ & $0.19^{+0.07}_{-0.08}$    &	$0.53^{+0.04}_{-0.02}$	&	$1.9^{+0.9}_{-1.0}$	& $1.73^{+0.35}_{-0.12}$	&	$1.69(0.18)$	& $1.08/281$		\\

Hatm($\times$2)	&	$16.7^{+8.2}_{-3.3}$	&	$2.56^{+0.06}_{-0.08}$	&	$0.17^{+0.03}_{-0.04}$	&	$1.00^{+0.03}_{-0.12}$	&	$1.3^{+1.0}_{-0.6}$	&	$0.42^{+0.06}_{-0.05}$	&	$4.9^{+7.9}_{-2.3}$	& $1.12^{+0.43}_{-0.31}$       & 	$3.54(0.11)$	&	$1.10/280$ \\
+BB+PL	&	$18.9^{+0.6}_{-0.6}$	&	$2.52^{+0.09}_{-0.09}$	&	$0.16^{+0.03}_{-0.03}$	&	$0.99^{+0.06}_{-0.22}$	&	$1.3^{+0.5}_{-0.7}$	&	$0.41^{+0.06}_{-0.04}$	&	$5.8^{+6.1}_{-4.8}$	&   $1.56$     & 	$3.58(0.13)$	&	$1.10/281$\\
	&	7	&	$2.56^{+0.06}_{-0.09}$	&	$0.17^{+0.05}_{-0.04}$	&	$1.04^{+0.12}_{-0.01}$	&	$1.1^{+1.1}_{-0.6}$	&	$0.48^{+0.02}_{-0.06}$	&	$2.1^{+2.2}_{-1.7}$	&   $1.15^{+0.43}_{-0.32}$     & 	$1.71(0.11)$	&	$1.10/281$\\
\hline
\colhead{} &  & \colhead{} & \colhead{} & \colhead{} & \colhead{} &  \colhead{$T_{\rm eff,3}$} & \colhead{$R_{\rm eff,3}$ }  & \colhead{$\tau$} & \colhead{} \\
\colhead{} &  & \colhead{} & \colhead{} & \colhead{} & \colhead{} &  \colhead{($10^6$ K)} & \colhead{(km)}  & \colhead{} & \colhead{} \\
\hline

CBB($\times$2)	&	$<$$0.1$	&	$2.63^{+0.03}_{-0.05}$	&	$0.08^{+0.03}_{-0.02}$	&	$0.93^{+0.02}_{-0.03}$	&	$0.55^{+0.21}_{-0.14}$	& -	 & -	& $0.115^{+0.006}_{-0.005}$ &	$1.33$	&	$1.31/283$	\\
CBB($\times$2)+BB	&	$19.0^{+9.0}_{-7.0}$	&	$3.01^{+0.09}_{-0.08}$	&	$0.05^{+0.02}_{-0.02}$	&	$1.45^{+0.08}_{-0.06}$	&	$0.25^{+0.13}_{-0.12}$	&	$0.43^{+0.06}_{-0.5}$	& $5.0^{+8.2}_{-3.2}$ & $0.102^{+0.007}_{-0.008}$ &	$2.74$	&	$1.07/281$	\\
	&	$7$	&	$3.03^{+0.11}_{-0.08}$	&	$0.05^{+0.02}_{-0.02}$	&	$1.49^{+0.10}_{-0.06}$	&	$0.22^{+0.09}_{-0.10}$	&	$0.52^{+0.04}_{-0.03}$	& $1.95^{+1.05}_{-1.07}$ & $0.109^{+0.007}_{-0.007}$ &	$1.68$	&	$1.08/281$
\enddata
\tablenotetext{a}{PL is a powerlaw, BB a blackbody, CBB a Comptonized blackbody, and Hatm a non-magnetic H
  atmosphere model. All uncertainties and limits quoted are at a 1$\sigma$ confidence level.}
\tablenotetext{b}{$R_{\rm eff}$ calculated assuming a distance of
  156.3 pc. For the H atmosphere model, the numbers quoted represent
  deprojected and redshift-corrected effective radii of one hot spot
  on a neutron star with $M=1.76$ M$_{\odot}$, $R=13.5$ km,
  $\alpha=36^{\circ}$, $\zeta=42.4^{\circ}$,
  $\Delta\alpha=-25^{\circ}$, and $\Delta\phi=-20^{\circ}$ (see text
  for definition of these parameters). The third, lowest temperature
  thermal component in all instances is a blackbody.}
\tablenotetext{c}{Unabsorbed X-ray flux (0.1--10 keV) in units of
  $10^{-13}$ ergs cm$^{-2}$ s$^{-1}$. The values in parentheses
  represent the flux contribution of the powerlaw component, where applicable.}

\end{deluxetable*}

\section{Phase-Averaged Spectroscopy}
\citet{Zavlin02} and \citet{Zavlin06} have found that the X-ray
spectrum of PSR J0437--4715 cannot be adequately represented by a
single or even most two component emission models. Based on this, in
the spectral fits I apply two plausible multi-component models: (i) a
multi-temperature thermal plus single non-thermal model, and (ii) a
multi-temperature Comptonized thermal model. For the thermal
components, I consider both a simple blackbody and a
non-magnetic H atmosphere spectrum \citep{Rom87,Zavlin96,McC04}, which
is appropriate for PSR J0437--4715 given its low magnetic field
($B\approx3\times10^8$ G). Even though polar cap heating in MSPs is
believed to be due to a return current of relativistic efrom from the
magnetosphere above the polar caps, the H atmosphere model assumption
of a heat source beneath the atmosphere is still fully valid.  This is
because the penetration depth of the impinging relativistic particles
from the magnetosphere is much greater ($\sim$2-3 orders of magnitude)
than the characteristic depth of the atmosphere, meaning that their
energy is deposited below the atmosphere \citep[see][for
  details]{Bog07}.

Blackbody fits to MSP spectra tend to produce inferred emission areas
that are substantially smaller (a factor of $\sim$10) than the
classical polar cap area, $R_{pc}=(2\pi R/cP)^{1/2}R$. In contrast, a
H atmosphere yields values comparable to $R_{pc}$ \citep{Beck02,
  Zavlin06, Bog09}.  It has been suggested that the discrepancy for
the blackbody model could be due to heating of only a small portion of
the polar caps \citep{Zha03}. More plausably, it suggests that an
atmospheric layer is present on the stellar surface, as expected from
evolutionary arguments. In particular, the peak intensity of a H
atmosphere continuum occurs at higher energies than a blackbody for
the same effective temperature. As a result, fitting a blackbody to H
atmosphere emission would produce a higher inferred temperature and,
as a result, a significantly smaller effective area.  Furthermore,
blackbody radiation has been shown to provide an inadequate
description of the thermal pulsations from this MSP
\citep{Zavlin98,Bog07}.  Nevertheless, for completeness and to enable
convenient comparison with previously published X-ray observations of
MSPs, I have considered a blackbody in the spectroscopic analysis as
well.

It is important to emphasize that even for phase-integrated
spectroscopy it is essential to properly account for the rotation of
the MSP and the geometric configuration of the hot spots, in order to
obtain reliable measurements of the temperatures and emission radii
\citep[][]{Psa00, Zavlin02,Bog07}. For this reason, I consider the
model of a rotating neutron star with two hot spots, based on the
prescription described in \citet{Belo02}, \citet{Pou03}, and
\citet{Vii04} but with the addition of a realistic H atmosphere
\citep[see][for further details]{Bog07}. This model incorporates the
angles between the spin and magnetic axes ($\alpha$) and the spin axis
and the line of sight ($\zeta$) as well as the mass and radius of the
pulsar as parameters in the spectral fits.  Thus, the derived
effective areas are deprojected and redshift-corrected. Moreover, the
inferred temperatures are as measured at the NS surface by correcting
them for gravitational redshift. For each polar cap I consider two
concentric emission regions -- a small high-temperature hot spot
surroundend by a cooler annular region, as implied by previous studies
\citep{Zavlin02,Zavlin06}.  In all instances, to calculate effective
radii, I consider the parallax distance of $156.3$ pc \citep{Del08}.
For the H atmosphere model, I assume a NS with $M=1.76$ M$_{\odot}$,
$R=13.5$ km. I also fix $\alpha=36^{\circ}$ and $\zeta=42^{\circ}$ and
account for the appearent offset of the secondary hot spot with the
parameters $\Delta\alpha=-25^{\circ}$ and $\Delta\phi=-20^{\circ}$, as
deduced from the pulse profile fits (see \S6). The various spectral
models, their best fit parameters and derived unabsorbed fluxes are
summarized in Table 1.  All uncertainties quoted are given at a
1$\sigma$ confidence level. Following the analysis by \citet{Dur12},
the fits are performed with all parameters free, as well as with a
fixed $N_{\rm H}=7\times10^{19}$ cm$^{-2}$ or a fixed
$\Gamma=1.56$. The latter was found by \citet{Dur12} to provide the
best fit to the broadband non-thermal spectrum of this pulsar.

Regardless of the spectral model employed, fitting the phase-averaged
MOS and pn datasets jointly results in statistically poor fits, with
typical $\chi_{\nu}\gtrsim 1.3$. In contrast, the fits to either the
MOS1/2 or pn data give $\chi_{\nu}\lesssim 1.1$ for the same assumed
models. This points to a possible discrepancy in the cross-calibration
of the detectors especially for the fast timing mode of the EPIC pn
instrument. Similar issues have been found in other studies using the
pn fast timing mode \citep{Zavlin06,Arch10}.  As a result, only the
MOS 1/2 data were considered for the phase-averaged fits as they span
a much wider energy range than the pn timing data.  The MOS1/2 data
exhibit significant narrow residuals at $\sim$1.7 keV . This feature
can be attributed to the Si-K edge\footnote{See
  \url{http://xmm2.esac.esa.int/docs/documents/CAL-TN-0018.pdf} for
  details.}.  Ignoring this interval results in significant
improvement in the quality of the fits.

\begin{deluxetable*}{lcccccccl}
\tabletypesize{\footnotesize} 
\tablecolumns{9} 
\tablewidth{0pc}
\tablecaption{EPIC pn results of phase-resolved spectroscopy for PSR J0437--4715.}
\tablehead{ \colhead{} & 
            \colhead{$N_{\rm H}$} &
            \colhead{$T_{\rm eff,1}$} & 
            \colhead{$R_{\rm eff,1}$\tablenotemark{b}} & 
            \colhead{$T_{\rm eff,2}$} & 
            \colhead{$R_{\rm eff,2}$\tablenotemark{b}} &  
            \colhead{$\Gamma$} & 
            \colhead{$F_{\rm X}$\tablenotemark{c}} & 
            \colhead{$\chi^2_{\nu}$/dof} \\
            \colhead{Model\tablenotemark{a}} & 
            \colhead{(cm$^{-2}$)} & 
            \colhead{($10^6$ K)} & 
            \colhead{(km)} & 
            \colhead{($10^6$ K)} & 
            \colhead{(km)} & 
            \colhead{} & 
            \colhead{(0.5--3 keV)}  &  }
\startdata
BB($\times$2)+PL	&	$<$$12.5$	&	$2.90^{+0.10}_{-0.11}$	&	$0.07^{+0.03}_{-0.03}$	&	$1.24^{+0.05}_{-0.06}$	&	$0.37^{+0.20}_{-0.18}$	&	$-0.14^{+1.38}_{-1.77}$ &	$9.4/6.0/4.9/6.4$	&	$1.02/480$	\\

	&	$<$$7.9$	&	$2.82^{+0.08}_{-0.04}$	&	$0.07^{+0.03}_{-0.03}$	&	$1.22^{+0.06}_{-0.06}$	&	$0.37^{+0.19}_{-0.19}$	&	$1.56$	&	$9.3/6.0/4.9/6.4$	&	$1.02/481$	\\

	&	$7$	&	$2.90^{+0.05}_{-0.06}$	&	$0.07^{+0.02}_{-0.02}$	&	$1.23^{+0.05}_{-0.06}$	&	$0.39^{+0.17}_{-0.14}$	&	$-0.11^{+1.67}_{-1.74}$	&	$9.6/6.1/5.0/6.6$	&	$1.02/481$	\\
\hline

Hatm($\times$2)+PL	&	$<$$4.3$	&	$2.22^{+0.04}_{-0.09}$	&	$0.16^{+0.07}_{-0.04}$	&	$0.55^{+0.05}_{-0.08}$	&	$6.0^{+6.0}_{-2.4}$	&	$2.35^{+0.17}_{-0.20}$ &	$9.3/6.1/4.9/6.2$	&	$1.06/486$	\\
	&	$<$$1.5$	&	$2.16^{+0.05}_{-0.07}$	&	$0.18^{+0.08}_{-0.06}$	&	$0.58^{+0.05}_{-0.05}$	&	$5.1^{+6.7}_{-3.6}$	&	$1.56$ &	$9.5/6.1/4.8/6.2$	&	$1.07/487$	\\
	&	$7$	&	$2.21^{+0.08}_{-0.10}$	&	$0.16^{+0.06}_{-0.05}$	&	$0.57^{+0.05}_{-0.08}$	&	$5.5^{+5.9}_{-3.5}$ &	$2.39^{+0.32}_{-0.18}$	&	$9.6/6.3 /5.0/6.3$	&	$1.06/487$

\enddata
\tablenotetext{a}{Model and parameter definitions are the same as in Table 1.}
\tablenotetext{b}{For the blackbody fits, the quoted effective radius
  is that measured at pulse maximum (corresponding to phase interval I
  in Figure 2). For the H atmosphere, the value is the true, 
  deprojected and redshift-corrected effective radius.}
\tablenotetext{c}{Unabsorbed X-ray fluxes (0.5--3 keV) in units of
  $10^{-13}$ ergs cm$^{-2}$ s$^{-1}$ for the four phase bins.}

\end{deluxetable*}

\subsection{Multi-Temperature Thermal Plus Powerlaw Spectrum}
Based on previous \textit{Chandra} and \textit{XMM-Newton}
observations, \citet{Zavlin02}, \citet{Zavlin06}, and \citet{Dur12}
have reported that the phase-integrated X-ray radiation from PSR
J0437--4715 is best represented by a two-temperature thermal plus a
single faint powerlaw components. When applied to the new EPIC MOS1/2
data, the same model is only marginally acceptable and shows
significant broad residuals at energies below $\sim$0.8 keV (middle
panels in Figure 3), similar to those observed in \textit{ROSAT} PSPC
data \citep[see][]{Zavlin02}. This suggest the presence of an
additional cool thermal component.  The atmosphere model used in this
study is only calculated down to $T_{\rm eff}=3\times10^5$ K since
below this temperature partial ionization becomes non-negligible
\citep{Zavlin96,McC04}. As a consequence, the third, cool thermal
component was represented by a blackbody. This addition results in a
significant improvement in the fit quality.

\citet{Dur12} have found that the spectral shape in the UV is
consistent with thermal emission from the bulk of the NS surface with
an effective temperature $(1.5-3.5)\times10^5$. The MOS 1/2 fits
suggest values in agreement with the upper bound of this range,
although the generally higher values may also indicate a temperature
gradient across the stellar surface, possibly as a function of
distance from the heated polar caps.

\subsection{Comptonized Thermal Spectrum}
As suggested in \citet{Bog06}, the power-law tail observed in the
spectrum of PSR J0437--4715 may arise due to inverse Compton
scattering (ICS) of the thermal X-ray radiation by relativistic
$e^{\pm}$ of small optical depth. It is plausible to assume that such
particles may be located in the magnetosphere above the polar caps.
For the spectroscopic analysis of PSR J0437--4715, I use the {\tt
  compbb} Comptonized blackbody model in {\tt XSPEC}
\citep{Nish86}. As additional parameters, this model includes a
scattering particle temperature $kT_e$ and optical depth $\tau$. For
the two thermal components the $\tau$ parameter was linked as,
presumably, both undergo scattering by the same $e^{\pm}$
population. The fit of a two-component Comptonized blackbody also
exhibits significant residuals at lower energies, which can be
accounted for by the additon of a cool blackbody component.

As expected, $kT_e$ and $\tau$ are strongly correlated, resulting in
acceptable fits over a wide range of the two parameters. Given that we
expect $kT_e$ to be much larger than $kT$, the particular choice of
$kT_e$ and the details of the energy distribution of the scattering
$e^{\pm}$ have little impact on the outcome of the X-ray spectral
fits.  Note that the inferred optical depth requires a particle
density in the magnetosphere (from the surface to the light cylinder
275 km above the surface) above the polar caps to be $\sim$$10^{4-6}$
times the Goldreich--Julian density.  It is interesting to note that
this value is comparable to that required for PSR J0737--3039A to
explain the observed radio eclipses in the double pulsar system
\citep{Lyut04,Arons05}.

%
%  FIGURE 4
%
\begin{figure}[t]
\begin{center}
\includegraphics[width=0.47\textwidth]{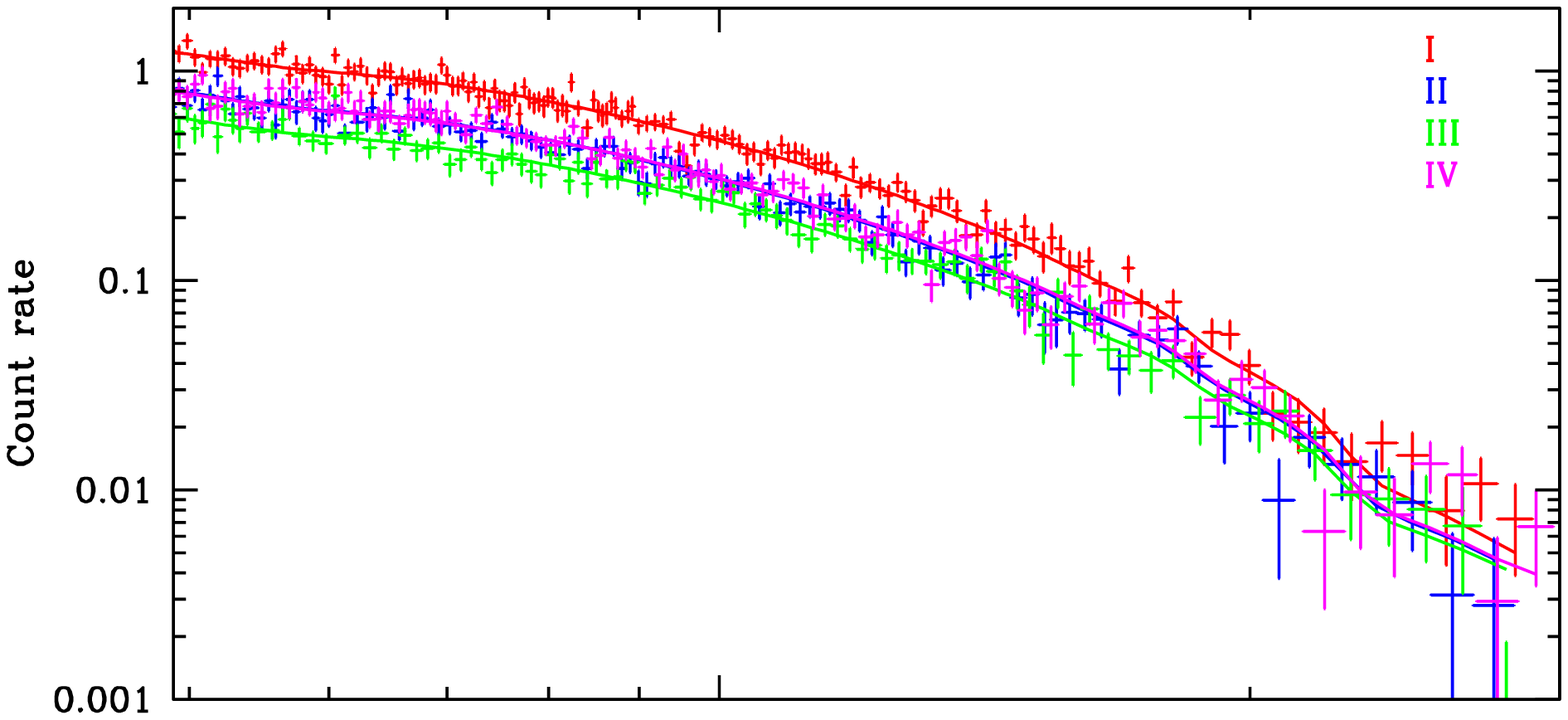}\\
\includegraphics[width=0.47\textwidth]{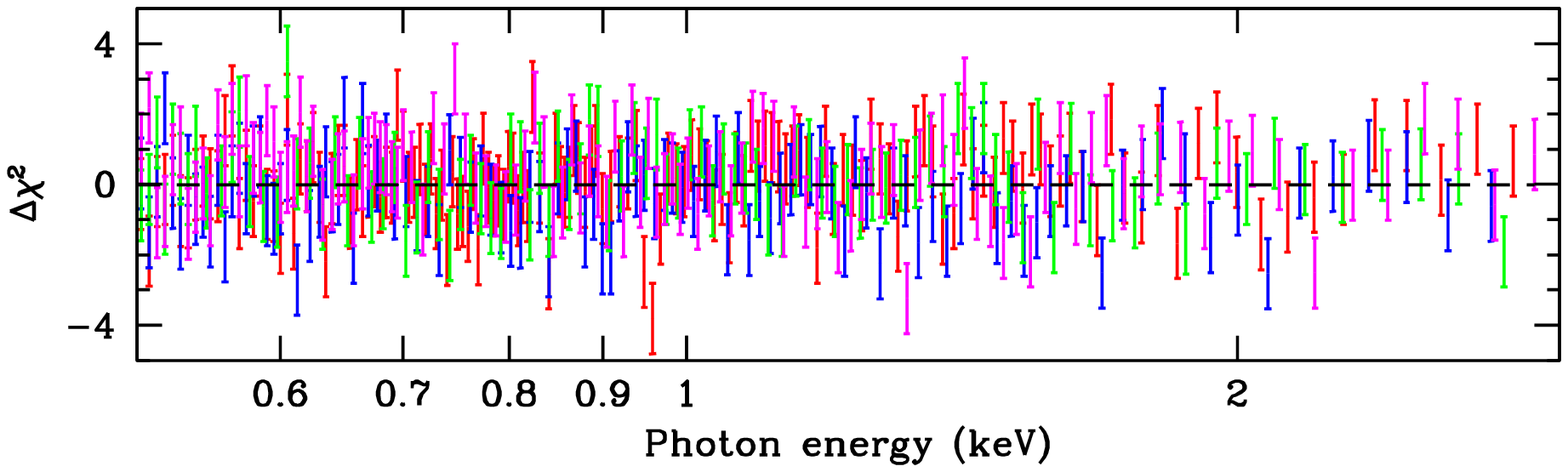}
\caption{The \textit{XMM-Newton} X-ray EPIC pn phase-resolved spectra
  of PSR J0437--4715 fitted with a H atmosphere model (\textit{solid
    lines}). The phase intervals used are shown in Figure 3. The
  bottom panel shows the fit residuals. The Roman numerals correspond
  to the phase intervals shown in Figure 2.}
\end{center}
\end{figure}

\section{Phase-Resolved Spectroscopy}
The sizable collection of source photons in the pn dataset permits an
analysis of the pulsar spectrum as a function of pulse phase. To this
end, I have divided the pn photons in the 0.5--3 keV range into four
phase intervals, with the first interval centered on the pulse peak
(see Figure 2). Due to the overwhealming background, the pn data above
$\sim$3 keV was excluded from the analysis. Moreover, as recommended
by the XMM-Newton Science Support Center, photons below 0.5 keV were
ignored in order to remove electronic noise artifacts seen in fast
timing mode.  Above this energy, the importance of the cool thermal
component seen in the MOS spectra is neglegible (as evident from the
bottom panel of Figure 3). Therefore, only two thermal components plus
the non-thermal component are used. This choice of energy cut also has
the consequence of making the fits nearly insensitive to $N_{\rm
  H}$. As evident from Table 2, the power-law component is also poorly
constrained since the $0.5-3$ keV band is strongly dominated by
thermal emission.

The spectral fits were performed on all four spectra simultaneously.
For the blackbody fits, the spectra were fit with the same
temperature, while allowing for the flux normalization to vary
independently for each spectrum.  In the case of the H atmosphere, all
parameters, including the flux normalization, were fitted jointly
among the four spectra since the pulse phase dependence of the
geometric projection of the emission area is taken into account
internally by the model for a given set of $M/R$, $\alpha$, and
$\zeta$. In other words, the flux normalization in this case
represents the true, deprojected area (as measured at the stellar
surface), which is the same at all phases. This was accomplished by
taking advantage of a functionality in {\tt XSPEC} that permits
simultaneous fitting of phase-resolved spectra\footnote{See
  \url{http://heasarc.nasa.gov/xanadu/xspec/manual/}
  for further details.}.  In addition to the anisotropic emission
pattern, an important distinction of the H atmosphere model from a
blackbody is the angular dependence of the emergent spectrum. In
particular, the continuum emission becomes softer for larger angles
with respect to the surface normal as a consequence of the temperature
gradient within the atmosphere. Observationally, this property should
appear as a shift in the peak of the spectrum to lower energies
\citep[see, e.g., Figure 2 in][]{Bog07}. Therefore, it is essential to
properly account for the geometry to ensure an accurate measurement of
the effective temperature of the radiation. Furthermore, since the
emission area is observed in projection it is important to correct for
this effect.

The results of the phase-resolved spectroscopic analysis are
summarized in Table 2. It is important to note that \citep[as shown in
  previous studies][]{Zavlin98,Bog07}, although the blackbody
model can reproduce the spectral shapes of this pulsar's
phase-resolved spectra for the same temperature, the implied
variations in the effective area cannot be explained by a rotating hot
spot model. This lends further support for anisotropic thermal
emission, which, in turn, implies the presence of an atmosphere on the
surface of PSR J0437--4715.

%
%    FIGURE 5
%
\begin{figure}[t]
\begin{center}
\includegraphics[width=0.45\textwidth]{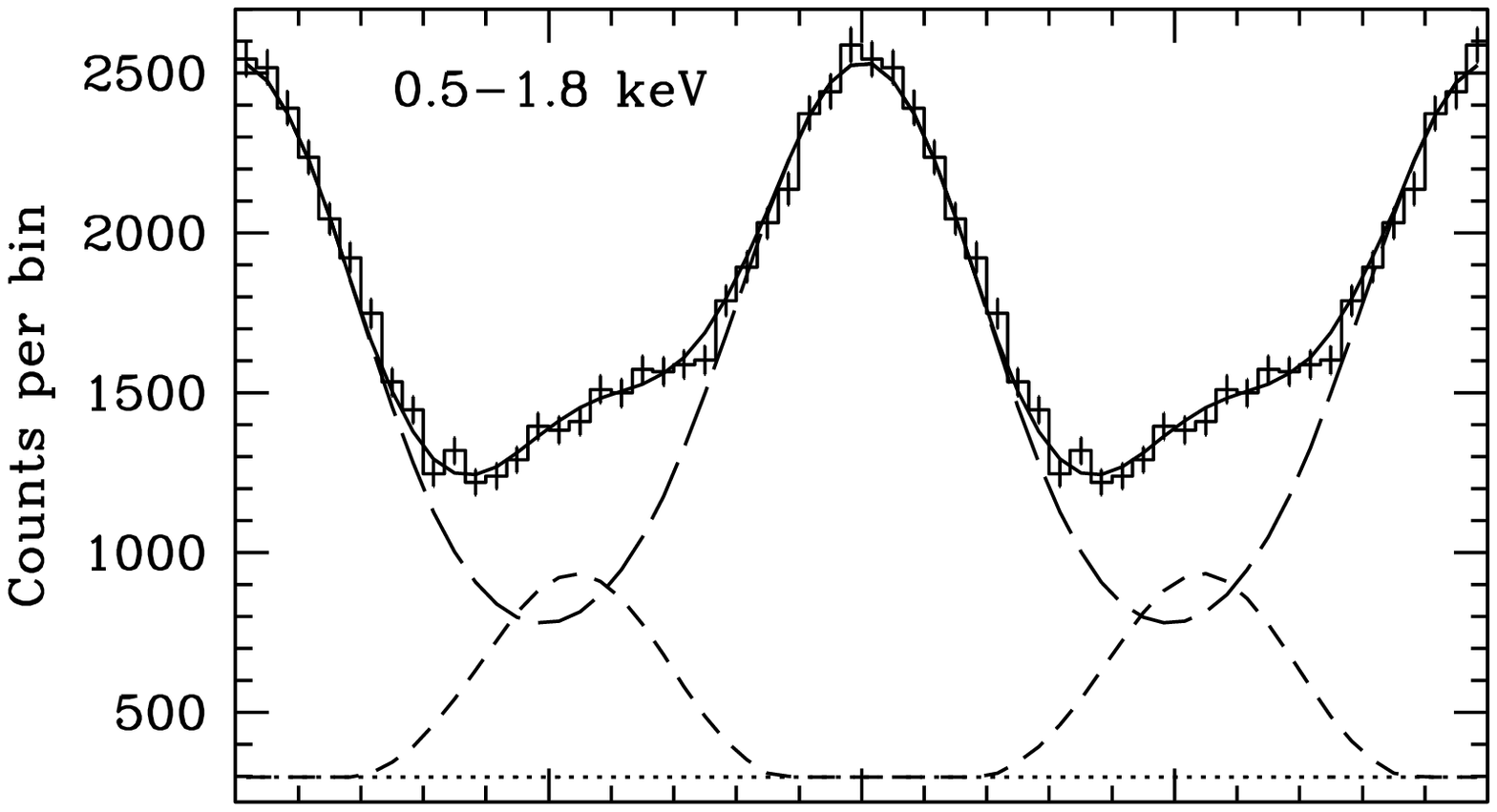}\\
\includegraphics[width=0.45\textwidth]{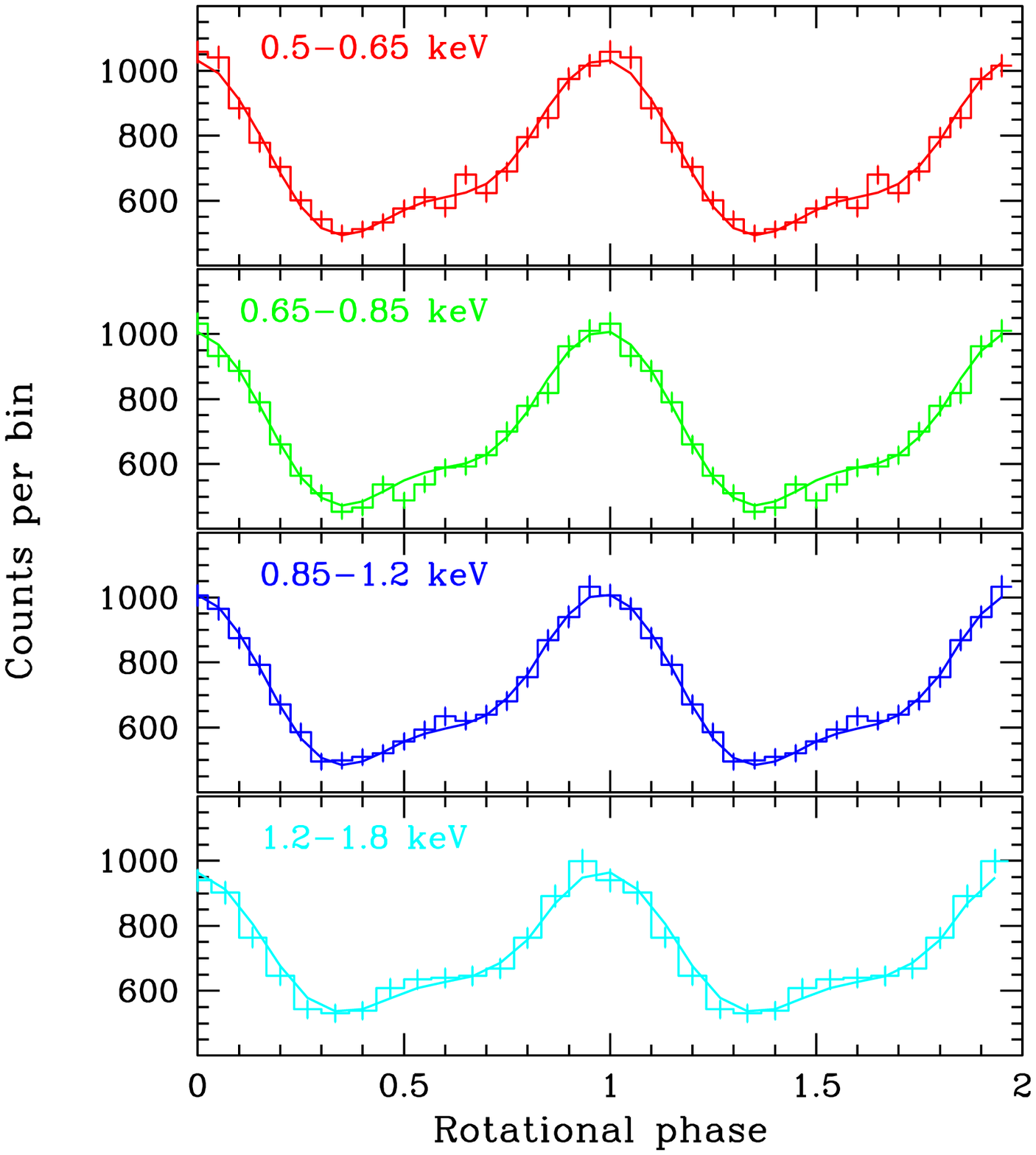}
\caption{(\textit{Top panel}) The \textit{XMM-Newton} EPIC pn pulse
  profile in the 0.5--1.8 keV range with the best fit model
  (\textit{solid line}). The individual contributions from each polar
  cap are shown with the dashed lines. The dotted line shows the
  background level. (\textit{Bottom four panels}) \textit{XMM-Newton}
  EPIC pn pulse profiles of PSR J0437--4715 in the 0.5--0.65,
  0.65--0.85, 0.85--1.2, and 1.2--1.8 keV bands (from top to bottom,
  respectively) fitted with a model of a rotating neutron star with
  two-temperature H atmosphere polar caps.  See text for best fit
  parameters.}
\end{center}
\end{figure}

%
%   FIGURE 6
%
\begin{figure}[t]
\begin{center}
\includegraphics[width=0.45\textwidth]{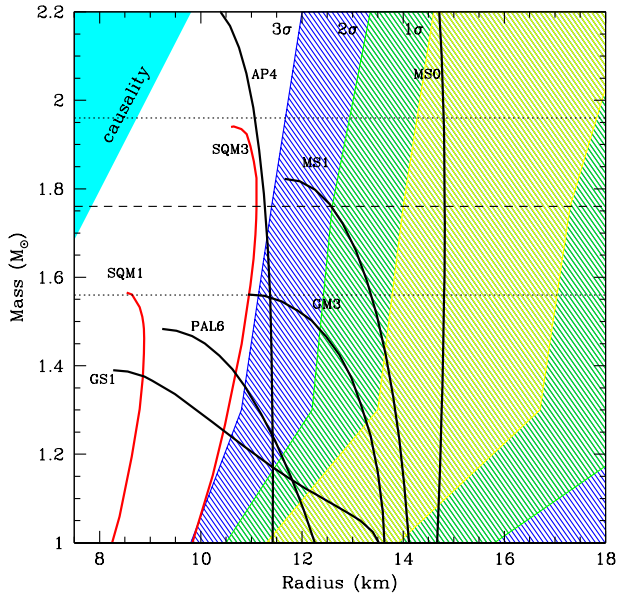}
\caption{The mass-radius plane for neutron stars showing the 1, 2, and
  3$\sigma$ confidence contours (\textit{yellow}, \textit{green}, and
  \textit{blue} hatched regions, respectively) for PSR
  J0437--4715. The solid lines are representative theoretical model
  tracks (from Lattimer \& Prakash 2004). The horizontal lines show
  the pulsar mass measurement from radio timing (\textit{dashed line})
  and the associated 1$\sigma$ uncertainties (\textit{dotted lines})
  from \citet{Verb08}. }
\end{center}
\end{figure}

\section{CONSTRAINTS ON THE NEUTRON STAR COMPACTNESS}
The observed pulsations of the surface thermal emission from PSR
J0437--4715 contain valuable information regarding the NS compactness,
which can be extracted via a constraint on the stellar mass-radius
relation.  To this end, I consider the model from \citet{Bog07}, which
assumes a rotating compact NS with two identical hot spots, each
corresponding to one of the magnetic polar caps.  As in the
spectroscopic analysis, each polar cap is assumed to consist of a
small high-temperature hot spot surroundend by a cooler annular
region.  The model considers a non-rotating Schwarzschild metric and
incorporates the relativistic Doppler effect and propagation time
delays using the prescription described in \citet{Belo02},
\citet{Pou03}, \citet{Vii04}, and \citet{Pou06}. This approach is
remarkably accurate as long as $P\gtrsim3$ ms \citep{Cad07,Mor07},
making it applicable to PSR J0437--4715. The NS surface is assumed to
be covered by a non-magnetic, optically-thick H atmosphere
\citep[][]{Rom87,Zavlin96,McC04}.  To allow direct comparison with the
data, the model was convolved with the EPIC pn plus thin filter
effective area and corrected for the encircled energy fraction and the
background level.

The X-ray pulse profile was fitted by considering ten parameters: the
two temperatures and effective radii of each hot spot ($T_1$, $T_2$,
$R_1$, and $R_2$), the angles $\alpha$ and $\zeta$, the NS radius $R$,
the displacement (in degrees) of the secondary hot spot from the
expected antipodal position ($\Delta \alpha$ and $\Delta \phi$), and
the phase $\phi$ of the main pulse.  High-precision radio timing
observations have measured the orbital inclination of the J0437--4715
binary system to be $i=137.6^{\circ}$ \citep{Verb08}. It is very
likely that the rotation axis of the MSP has aligned with the orbital
angular momentum vector as the NS was spun-up during the low-mass
X-ray binary stage of its evolution. This scenario implies
$\zeta$$\approx137.6^{\circ}$ or, equivalently,
$\zeta\approx42.4^{\circ}$. Therefore, I fix the value of $\zeta$ to
the latter value.  The H absorption column along the line of sight was
set to $N_{\rm H}=7 \times 10^{19}$ cm$^{-2}$, while the distance was
fixed at $156.3$ pc \citep{Del08}. For all intents and purposes the
0.8\% uncertainty in the parallax distance is completely
negligible. Varying $N_{\rm H}$ over the range $(1-10)\times10^{19}$
cm$^{-2}$ results in virtually no change in the fit since its effect
is unimportant above 0.5 keV. In the analysis, I assumed a fixed mass
and allowed $R$ to vary. The procedure was repeated for $M=0.7$,
$1.0$, $1.3$, $1.76$ and $2.2$ M$_{\odot}$.  The fit was conducted
simultaneously for four photon energy ranges ($0.5-0.65$ keV,
$0.65-0.85$ keV, $0.85-1.2$ keV, and $1.2-1.8$ keV) in order to weaken
any covariance between the parameters that define the emission
spectrum (temperatures and effectve radii) from the geometric and
relativistic parameters, which do not depend on energy. The energy
ranges were chosen in such a way as to yield comparable numbers of
photons for all bands and to reduce the contribution of background
photons.  To include the non-thermal component seen in the total
spectrum, I absorbed the expected number of photons per bin into the
uncertainties, assuming an unpulsed contribution and a $\Gamma=1.56$
powerlaw spectrum. The best fit values for each parameter were
determined by searching the $\chi^2$ space. To ensure that the these
values are the true absolute minima of the parameter space, the
procedure was repeated for a wide range of initial values for each
parameter.  Due to the complicated topology of the parameter space,
the confidence intervals of the best fit parameters were determined by
way of Monte Carlo simulations, using $5\times10^3$ realizations for
each of the four assumed NS masses.

The best fits to the X-ray pulsations for the H atmosphere hot spot
model are shown in Figure 5. The plotted best fit yields
$\chi^2_{\nu}=0.9$ for 65 degrees of freedom. As seen previously,
unlike a blackbody model, H atmosphere emission can account for the
morphology of the X-ray pulsations and large amplitude remarkably
well. In comparison with the modeling in \citet{Bog07}, the best fit
shown in Figure 5 confirms the upward inflection in the lightcurve
around phases $0.5-0.75$, which appears to arise due to the secondary
hot spot.  Figure 6 shows the $M-R$ plane with representative models
for different varieties of NS EOS \citep[see][]{Latt01} and the limits
derived for PSR J0437--4715, with the hatched areas representing the
$1$, $2$, and $3$$\sigma$ confidence regions. The NS radius is
constrained to be $>11.1$ km at 3$\sigma$ confidence, assuming
$M=1.76$ M$_{\odot}$ for all combinations of the other parameters.  If
taken in combination with the best available mass measurement from
radio timing \citep{Verb08}, the limits are inconsistent with all but
the stiffest equations of state. It is interesting to note that the
lower bounds on the confidence intervals are much tighter. This occurs
because for more compact stars it is not possible to produce the large
observed pulse amplitude, owing to the stronger gravitational bending
of light effect, which acts to surpress rotation-induced
modulations. In contrast, the upper bounds allow for very large stars
since large-amplitude pulsations can be produced more easily for a
wide range of viewing geometries.

The fits to the X-ray pulsations also yield constraints on the
magnetic field configuration of PSR J0437--4715. In particular,
assuming $M=1.76$ M$_{\odot}$, the magnetic inclination is
constrainted to be $\alpha=36^{+4}_{-3}$ degrees (at
1$\sigma$ confidence), suggesting a significant misalignment between
the magnetic and rotation axes\footnote{By convention, $\alpha$ is
  reckoned from the spin pole towards the equator.}. The latitudinal
offet of the secondary polar cap is constrained to be $\Delta \alpha=
-25^{+6}_{-4}$ degrees. As defined, a negative
$\Delta\alpha$ corresponds to a northward shift from the antipodal
position. The displacement in the longitudinal direction is found to
be $\Delta\phi=-20^{+2}_{-1}$ degrees, indicating that
the secondary spot trails the antipodal position with respect to the
sense of rotation, as evident from the phase lag of the secondary
pulse in Figure 5. Based on Equations 2 and 3 in \citet{Bog08} and
assuming the nominal best fit NS radius of 13.5 km, the corresponding
net offset of the secondary hot spot from the antipodal position
across the NS surface is $\sim$6 km, while the implied displacement of
the magnetic dipole axis from the stellar center is $\Delta x\approx3$
km.

\section{CONCLUSION}

In this paper, I have presented an analysis of the deepest X-ray
observation of a radio MSP to date, namely, an \textit{XMM-Newton}
exposure of PSR J0437--4715.  The data permit a significantly better
characterization of the spectral continuum of this pulsar in the
0.1--10 keV band compared to previous studies. Although the observed
radiation from J0437--4715 can be described well by multiple models,
in all cases the vast majority of soft photons ($\gtrsim$90\%) appear
to be thermal in nature. This thermal radiation exhibits at least
three components, with the hottest two having total effective areas
consistent with the expected polar cap size. The coolest component, on
the other hand, appears to cover a significant portion of the stellar
surface, consistent with findings based on UV observations
\citep{Kar04,Dur12}. Since passive cooling of any primordial heat
stored in the NS cannot account for the high global NS temperature
($\gtrsim$$10^5$ K), various heating mechanisms have been envoked to
explain the excess, including: re-radiation of the energy deposited in
the inner crust by polar cap heating, rotochemical heating
\citep{Fern05}, ``gravitochemical'' heating caused by a time-varying
gravitational constant \citep{Jof06}, or capture of exotic particles
in the NS interior \citep{Han02}.  See \citet{Kar04} and \citet{Dur12}
for further discussions of the various possible heating mechanisms
that may operate in PSR J0437--4715.

An important goal of future observations will be to establish the true
nature of the hard photons ($\gtrsim3$ keV) from PSR J0437--4715. This
can be best accomplished through detailed phase-resolved
spectroscopy. Better characterization of the optical and $\gamma$-ray
emission from this pulsar may offer additional clues into the nature
of this radiation.  In addition, with the recent launch of
NuSTAR\footnote{http://www.nustar.caltech.edu/home}, a hard X-ray
focusing telescope covering the 6--79 keV range, further insight could
be gained into the connection between the harder X-ray and $\gamma$-ray
emission from this pulsar.

For the first time, I conduct phase-resolved X-ray spectroscopy of a
radio pulsar that correctly accounts for the system geometry and the
radiation pattern of the atmosphere by jointly fitting the spectra
from all phase intervals. With greatly increased photon statistics,
this potentially powerful approach can yield stringent constraints on
the NS parameters by utilizing a two-dimensional approach to
spectroscopy in the photon energy--pulse phase plane.

The model pulse profiles in Figure 5 imply a significant offset of the
secondary hot spot from the antipodal position, which in turn, suggest
deviation from the standard centered magnetic dipole configuration.
As demonstrated by \citet{Hard11}, even a slight azimuthally
asymmetric displacement of the pulsar magnetic field can substantially
amplify the accelerating electric field on one side of the polar cap,
which when combined with a smaller field line radius of curvature,
leads to larger pair multiplicity. As a consequence, the death line
for producing $e^{\mp}$ by curvature radiation shifts downward in the
period-period derivative diagram, resulting in a larger number of
pulsars having high pair multiplicities. If common among MSPs, and
pulsars in general, these magnetic field distortions could have
profound implications for studies of the pulsar population, modelling
high energy pulsed radiation, and constraining the pulsar contribution
to cosmic ray positrons.

The analysis presented herein confirms that the X-ray pulsed fraction
od PSR J0437--4715 is consistent with the existence of an atmospheric
layer on the NS surface and cannot be reproduced by a blackbody.
Modelling of the thermal pulsations gives a limit on the NS radius of
$>11.1$ km (3$\sigma$ conf.) assuming a mass of $1.76$
M$_{\odot}$. This lower bound is much stricter compared to the limit
obtained from the shallower archival observations.  This value is in
agreement with several NS radius measurements using quiescent low-mass
X-ray binaries \citep[e.g.,][]{Rut01,Heinke06,Webb07} and X-ray bursts
\citep[see, e.g.,][]{Ozel06,Ozel10}.  However, it is important to
emphasize that these methods have resulted in a wide range of often
contradictory constraints on the NS radius \citep[see, e.g.,][for the
  case of X-ray bursts]{Sul11,Ozel12}. This highlights the clear need
to investigate the systematic uncertainties and reliability of these
model-dependent approaches to constraining NS structure, including the
method employed in this present paper.

This issue notwithstanding, as demonstrated in \citet{Bog08}, deeper
X-ray observations of PSR J0437--4715 should result in even stricter
constraints on $M/R$, especially when combined with further refinement
in the NS mass measurement from radio timing and joint modelling of
the pulsed X-ray and $\gamma$-ray emission \citep[see,
  e.g.,][]{Vent09} to better constrain the system geometry. Such
efforts may produce definitive measurements of the $M-R$ relation, and
by extension, the state of cold, dense matter.

\acknowledgements I thank Z.~Arzoumanian and V.~Kaspi for numerous
insightful discussions.  This research was supported in part by a
Canadian Institute for Advanced Research Junior Fellowship.  The work
presented was based on observations obtained with \textit{XMM-Newton},
an ESA science mission with instruments and contributions directly
funded by ESA Member States and NASA. The research in this paper has
made use of the NASA Astrophysics Data System (ADS).\\

Facilities: \textit{XMM-Newton} (EPIC)
%\facilities{XMM-Newton}

\end{document}